\documentstyle[12pt,aaspp4]{article}

\lefthead{Secker, Harris and Plummer}
\righthead{Dwarf Galaxies In Coma. II.}

\begin{document}

\title{Dwarf Galaxies in the Coma Cluster. II. Photometry and Analysis}

\author{Jeff Secker\\
Program in Astronomy, Washington State University,
Pullman, WA 99164-3113 USA\\
{\em secker@delta.math.wsu.edu}}

\author{William E. Harris\altaffilmark{1}\\
Department of Physics and Astronomy, McMaster University,
Hamilton, Ontario L8S 4M1 Canada\\
{\em harris@physics.mcmaster.ca}}

\author{Julia D. Plummer\altaffilmark{2}\\
Program in Astronomy, Washington State University,
Pullman, WA 99164-3113 USA}

\altaffiltext{1}{Visiting Astronomer, Kitt Peak National Observatory, 
operated by AURA, Inc.\ under contract to the National Science Foundation.} 
\altaffiltext{2}{Current address: Department of Astronomy, University of 
Michigan, 830 Dennison, 501 East University Avenue, Ann Arbor, MI 
48109-1090; {\em plummer@astro.lsa.umich.edu}}

\begin{abstract}
We use the data set derived in our previous paper (Secker \& Harris
1997) to study the dwarf galaxy population in the central $\simeq 700$
arcmin$^2$ of the Coma cluster, the majority of which are early-type
dwarf elliptical (dE) galaxies.  Analysis of the
statistically-decontaminated dE galaxy sequence in the color-magnitude
diagram reveals that the mean dE color at $R = 18.0$ mag is $(B-R)
\simeq 1.4$ mag, but that a highly significant trend of color with
magnitude exists ($\Delta (B-R)/\Delta R = -0.056\pm0.002$ mag) in the
sense that fainter dEs are bluer and thus presumably more metal-poor.
The mean color of the faintest dEs in our sample is $(B-R) \simeq
1.15$ mag, consistent with a color measurement of the diffuse
intracluster light in the Coma core.  This intracluster light could
then have originated from the tidal disruption of faint dEs in the
cluster core.  The total galaxy luminosity function (LF) is well
modeled as the sum of a log-normal distribution for the giant
galaxies, and a Schechter function for the dE galaxies with a
faint-end slope $\alpha = -1.41\pm0.05$. This value of $\alpha$ is
consistent with those measured for the Virgo and Fornax clusters.  The
spatial distribution of the faint dE galaxies ($19.0 < R \le 22.5$
mag) is well fit by a standard King model with a central surface
density of $\Sigma_0 = 1.17$ dEs arcmin$^{-2}$ and a core radius
$R_{\rm c} = 22.15$ arcmin ($\simeq 0.46h^{-1}$ Mpc).  This core is
significantly larger than the $R_{\rm c} = 13.71$ arcmin ($\simeq
0.29h^{-1}$ Mpc) found for the cluster giants and the brighter dEs ($R
\le 19.0$ mag), again consistent with the idea that faint dEs in the
dense core have been disrupted.  Finally, we find that most dEs belong
to the general Coma cluster potential rather than as satellites of
individual giant galaxies: An analysis of the number counts around 10
cluster giants reveals that they each have on average $4\pm 1$ dE
companions within a projected radius of $13.9h^{-1}$ kpc.
\end{abstract}


\section{INTRODUCTION} 

Like their giant-galaxy counterparts, dwarf galaxies are divided into
{\em early-} and {\em late-} types.  Late-type dwarfs are found
predominately in the field or the less dense regions of galaxy
clusters, and are dominated in number by the dwarf irregular (dIr)
galaxies, with a small contribution from dwarf spirals (Schombert et
al. 1995).  By contrast, dwarf ellipticals (dEs) are found in large
numbers within rather dense, populous cluster environments.  In this
paper, we present a study of the properties of the dwarfs in the Coma
cluster core, which are almost entirely of the dE type.

The prototypical dE is a smooth, centrally concentrated, low surface
brightness galaxy.  The majority of more luminous dE galaxies are
nucleated (dE,N), in which a central spike contains up to 20 percent
of the total luminosity of the galaxy (van den Bergh 1986; Ferguson \&
Binggeli 1994; Durrell 1997).  These nuclei are most likely
supermassive star clusters with typical radii $\lesssim 50$ pc (hence
unresolved at the distance of Coma with ground-based resolution), and
their integrated colors are usually indistinguishable from the
surrounding galaxy (Caldwell \& Bothun 1987).  The faintest dE
galaxies, which have a less than average central concentration, are
often referred to as dwarf spheroidals (dSph); see Gallagher \& Wyse
(1994).  This distinction is simply one of total magnitude, since the
characteristics of dE galaxies (see below) carry over to the fainter
dSph galaxies.  For about 25 dwarf members of the Virgo cluster,
Sandage \& Binggeli (1984) introduced the dS0 class.  These rare
objects are intended to be analogous to the regular S0 galaxies, and
show evidence for a two-component structure (i.e., a core plus
flattened disk).  The dividing lines among all these various
subclasses are approximate and can even differ significantly between
individual studies.  In our analysis to follow, we make no effort to
separate the sample of dE galaxies into their morphological groups,
simply because our image resolution is insufficient to do so.  Thus in
this paper, we define dE galaxies empirically as low luminosity, low
surface brightness galaxies in the range $-19.5 \lesssim M_R \lesssim
-9.5$ mag, with colors typically near $1.3 \lesssim (B-R) \lesssim
1.5$ mag.

The great majority of dE galaxies are located in the dense environment
of rich clusters of galaxies.  At the present time, the formation of
dE galaxies in rich clusters is not well understood (Ferguson \&
Binggeli 1994; Moore et al. 1996; Secker 1996), and so we are unsure
of their age and stellar composition.  While most are thought to be
composed of a metal-poor stellar population similar in age to globular
clusters (Ferguson \& Binggeli 1994), some unknown fraction will most
likely be similar to the Local Group dE and dSph galaxies, many of
which show evidence for more than one episode of star formation
(Caldwell \& Bothun 1987; Smecker-Hane et al. 1994; Sarajedini \&
Layden 1995).  What appears to be known is that dEs form a population
which is fundamentally different from regular elliptical (E) galaxies
(although see Jerjen \& Binggeli 1997):

(a) dE galaxies in general become less luminous with {\em decreasing}
central surface brightness, contrary to the class of regular E
galaxies for which the core radius and central surface brightness {\em
increases} toward fainter magnitude (Kormendy 1977; Caldwell \& Bothun
1987; Ferguson \& Binggeli 1994). These features provide the basis for
visual discrimination between dE and E galaxies near the same total
magnitude: the dE galaxies are those with the lower surface
brightness.

(b) dEs are characterized by smooth (as compared to dIr) surface
brightness profiles, with the non-nucleated dEs being well fit by a
single exponential (Ichikawa et al. 1986; Ferguson \& Binggeli 1994)
or a modified-exponential profile (Cellone, Forte \& Geisler 1994;
Jerjen \& Binggeli 1997).  This differs from E galaxies, whose light
distribution is well described by a de Vaucouleurs $R^{1/4}$ law.
Thus in principle, the bright dE galaxies can be distinguished from
faint E galaxies (at the same magnitude) on the basis of their surface
brightness profiles.

(c) while {\em color gradients} are observed in some dE galaxies,
there does not appear to be any preference for red or blue color
gradients (Durrell 1997).  This differs from the giant ellipticals,
which generally become bluer (a metallicity effect) at larger radius.

(d) dEs appear to be dominated by extensive and extremely massive
dark-matter halos.  Measurements of the central stellar velocity
dispersions indicate that $M/L \propto L^{0.2}$ for Es, but that $M/L
\propto L^{-0.4}$ for dEs (Ferguson \& Binggeli 1994). Typical
values seem to range from $M/L = 4$ or 5 $M_{\odot}/L_{\odot}$ for
bright dEs in Virgo, up to $M/L \simeq 100$ for Draco, a dSph in the
Local Group (Bender, Paquet \& Nieto 1991; Peterson \& Caldwell
1993). This observed trend in $M/L$ for dEs is consistent with the
models of Dekel \& Silk (1986): for the case of dominant dark-matter
halos, they predict a mass-to-light ratio which varies as $L^{-0.37}$.
That is, the efficiency in which gas is converted into stars decreases
with luminosity for dE galaxies, as the lower-mass dwarfs lose a
larger fraction of their gas.

There have been three recent photometric studies of the small-galaxy
population in Coma, each of which has a significant overlap with ours:
(1) Thompson \& Gregory (1993; extending the study of Thompson \&
Gregory 1980) is based upon photographic $m_b$ magnitudes and $(b-r)$
colors, complete to a limiting magnitude of $m_b = 20.0$ mag (i.e., $R
\simeq 18.6$ mag). Their survey covers an area of approximately four
square degrees, and provides a thorough analysis of many aspects of
the Coma dwarf galaxy population. (2) Biviano et al. (1995) utilize a
combination of redshifts and $(b-r)$ colors, together with
photographic magnitudes, to obtain a sample of cluster members
complete to a limit of $m_b = 20.0$ mag.  With this data set they
complete a statistical analysis of the galaxy luminosity function for
an area of the cluster core 1260 arcmin$^2$, centered upon NGC 4878
and NGC 4889. (3) Bernstein et al. (1995) analyze deep $R$-band CCD
images taken with the KPNO 4-meter telescope in $1.3^{''}$ seeing.
Their single field is 56 arcmin$^2$, located to the south of both NGC
4889 and NGC 4874.  They perform number counts of all objects to a
limit of $R = 25.5$ mag, with an excellent job of statistical number
correction using five randomly-selected control fields.

The primary difference between these studies and ours is that we
combine a moderately large spatial coverage ($\simeq 700$ arcmin$^2$)
with accurate $R$- and $B$-band CCD photometry and a moderately deep
limiting magnitude.  Our cluster membership is based upon the object's
$(B-R)$ color, and as illustrated by Biviano et al. (1995; a
comparison of redshift-selected and color-selected cluster
definitions), this technique results in a minimal contamination from
background galaxies.  As described in our companion Paper I (Secker \&
Harris 1997), our CCD images cover a large fraction of the core
environment of the Coma cluster, i.e. the region surrounding the
supergiants NGC 4874 and NGC 4889, with another field extending
southward to 23 arcmin from NGC 4874. These images provide an
extremely large sample of candidate dwarf galaxies, the vast majority
of which are the early-type dEs, and our analysis of this sample
yields significant correlations relevant to their formation and
subsequent dynamical evolution in this ultradense environment.

In Paper I we describe the observations and data reduction techniques
for a sample of candidate cluster galaxies. The dES are clearly
evident in the color-magnitude diagram (CMD) as a tightly defined
sequence of objects not present on the background control field.  In
this paper we analyze several features of our database for the dE
galaxy population: In Section 2, we discuss sample definition,
photometric calibration and magnitude completeness.  In Section 3, we
analyze the dependence of dE color on integrated magnitude, using the
control-field subtracted color-magnitude diagram.  In Section 4 we
compute the faint-end slope of the Schechter function, and model the
net galaxy luminosity function, decomposing it into two components.
In Section 5, we analyze and compare control-field corrected radial
number-density profiles for magnitude-selected subsamples of the
cluster galaxies. Finally, in Section 6 we model and subtract a subset
of the cluster giants, in an effort to quantify the number of bound
companions in this dense environment.

\section{Sample Definition, Calibration and Completeness}

\subsection{Calibration Procedure}

The calibration of our $B$- and $R$-band magnitude zeropoints and
other coefficients was accomplished with repeated observations of
four standard star fields: M67 (Schild 1983), M92, NGC 2419 and NGC
4147 (Davis 1990), summarized in Paper I.  To our $B$- and $R$-
standard star measurements we fit standard linear photometric
transfer equations

\begin{eqnarray}
R     & = & r - ZP_{R} - a_{1}X_R - a_{2}(B-R)             \nonumber \\
(B-R) & = & \frac{\left[ (b-ZP_B)-(r-ZP_R)\right] -a_{3}X_B + a_{1}X_R}
{1-(a_4-a_2)}
\label{eq:cal2}
\end{eqnarray}

\noindent where $b$ and $r$ represent instrumental aperture magnitudes of
the standard stars, $ZP_{B}$ and $ZP_{R}$ denote the magnitude
zeropoint terms, $a_{1}$ and $a_{3}$ represent the extinction
coefficients, and $a_{2}$ and $a_{4}$ represent the color
coefficients.  The $b$ and $r$ magnitudes were derived with {\sc
DAOPHOT} (Stetson 1987), by summing the pixel intensity within a
10-pixel aperture radius, subtracting the mode of the sky
distribution computed within a sky annulus of inner radius 20
pixels and width of five pixels, and normalizing to a 1 second
exposure.

We assume that the color coefficients are constant from night to night
for a given filter, whereas the zeropoint term and the airmass
coefficient can vary between nights.  For both nights, the
coefficients of these transformation equations were determined using
the simultaneous multilinear least squares reduction method developed
by Harris et al. (1981).  To best constrain our solutions, we adopted
typical values for the extinction coefficients of $a_{1}$ = 0.130 and
$a_{3}$ = 0.283 (Landolt 1982).  Reducing both nights of data
simultaneously (but $R$ and $B$ separately) yielded the zeropoints and
coefficients given in Table \ref{tbl1}.  With the mean airmass values
tabulated in Paper I, all quantities on the right-hand side of
(\ref{eq:cal2}) are known, and our magnitudes and colors are
completely calibrated.

\placetable{tbl1}

\subsection{Magnitude Completeness and Photometry Comparison}

The completeness of detection as a function of magnitude, $f(m)$, was
first estimated by matching object detection lists from overlap
regions for the three program-field $R$-band master images.  There are
two of these overlap regions: the North-South region, consisting of
$280\times 2000$ px$^2$ in common between the {\sc NGC 4874} and {\sc
NGC 4889} fields, and the East-West region, consisting of $1500\times
450$ px$^2$ in common between the {\sc NGC 4874} and the {\sc NGC 4874
South} fields. These regions extend from the strong light gradients
between {\sc NGC 4874} and {\sc NGC 4889} to the relatively flat sky
levels south of {\sc NGC 4874} and thus provide a sensible overall
average estimate for $f(m)$.  As will be seen in more detail below, $f
< 0.8$ above our adopted limiting magnitude of $R = 22.5$ mag; thus
there is less than a four percent chance that a real object will be
missed on both images on an overlap region.  An advantage to this
method (as opposed to artificial star simulations; see Section 5) is
that the objects we compare are real; they span a large range in
surface brightness, and they have been through the complete range of
image preprocessing, detection and measurement.

Consider first the East-West overlap region.  We define $N_T(m)$ as
the total (combined) number of objects detected on both the East and
the West sections.  Then, we define $N_m(m)$ as the number of matched
objects; that is, the subset of objects detected on the West ({\sc NGC
4874}) image, and detected (measured at any magnitude) on the
corresponding East overlap region.  For the purpose of binning, the
object's ``true'' magnitude is adopted to be that measured on the {\sc
NGC 4874} master image.  Then the fractional completeness in each
0.5-magnitude bin is given by

\begin{equation}
f(m) = \frac{N_m(m)}{N_T(m) - N_m(m)},
\label{eq:fraccomp}
\end{equation}

\noindent and we adopt Bolte's (1989) expression for the uncertainty 
in $f(m)$, given by

\begin{equation}
\sigma(f) = \frac{f(1-f)}{N_T}.
\label{eq:comperr}
\end{equation}

\noindent Results of this matching procedure from the two separate 
overlap regions were combined together to yield an averaged
completeness function, for which the effects of small-number
statistics are somewhat reduced.

In Figure \ref{MATCHEDcompfcn} we plot $f(m)$ over the magnitude range
of interest, derived using (\ref{eq:fraccomp}) and (\ref{eq:comperr}).
The {\em open circles} denote the completeness function derived using
{\em all} measured objects, while the {\em solid circles} denote the
completeness function derived using the subsample of objects within
the color range $0.7 \le (B-R) \le 1.9$ mag (Section 2.3).  While
these two functions are consistent within their uncertainty ranges,
the color-restricted subsample is consistently more complete at all
magnitudes.  It is the $f(m)$ defined by this color-restricted sample
which we adopt for statistical corrections in Section 4.  Based upon
an inspection of this completeness plot, and the behavior of the
photometric error with magnitude (Paper I), we adopt a limiting
magnitude of $R = 22.5$ mag, and discard all objects with total
magnitudes below this limit.  Note that the object sample used to
define these functions was truncated at $R = 23.5$.  Since the scatter
in the total magnitude is about 0.1-mag at $R = 22.5$ mag, the faint
end of our derived $f(m)$ is not affected by this truncation.

\placefigure{MATCHEDcompfcn}

In the range $18 \le R \le 22.5$
mag, the color restricted subsample of objects is well represented by
a straight line, as plotted in Figure \ref{MATCHEDcompfcn} (dashed
line): 

\begin{equation}
f(m) = \left\{
\begin{array}{rr}
-R/22.5 + 1.8; & 18 \le R \le 22.5 {\rm \ \ mag} \\
          1.0; &        R < 18.0   {\rm \ \ mag} \\
\end{array}
\right.
\label{eq:compfcn}
\end{equation}

\noindent In addition to the magnitude incompleteness, incompleteness 
can also occur in color and spatial distributions.  Brighter than $R = 22.5$
mag, we were able to measure $B$-band magnitudes for virtually all
detected objects.  (We discarded a handful of unmeasured objects, but
well less than one percent.)  Thus, {\em we are essentially 100
percent complete for the color measurements}.  Concerning spatial
incompleteness, we discard a circular region of radius 150 pixels
centered on both of the two supergiant galaxies, and a circular region
of radius 100 pixels centered on the bright saturated star located
north of {\sc NGC 4874}.  On the original master images, the light
gradients in these regions were too extreme to permit adequate
filtering and useful photometry.  To a lesser extent, all of the giant
(bright and extended) cluster galaxies contribute to a small (but
non-negligible) degree of spatial incompleteness, as we cannot measure
a small, faint, superimposed object using aperture photometry (Secker
\& Harris 1996).

The set of all objects detected on both (i) an overlap region of the
{\sc NGC 4874} image and on (ii) another of the program fields, can be
used to compare total magnitudes and colors as a function of apparent
total $R$ magnitude.  In the top panel of Figure \ref{OVERLAPcompare}
we plot the magnitude difference $\Delta R$ mag versus the total
magnitude, ranging from our completeness limit up to $R = 14.5$ mag.
On this figure, $\Delta R$ is defined as the {\sc NGC 4874} magnitude
minus the other measured magnitude; the cross symbols denote the 416
objects matched with the {\sc NGC 4889} field, and the solid circles
denote the 448 objects matched with the {\sc NGC 4874 South} field.
It is immediately obvious that our magnitude scale is consistent
between fields: the vast majority of the objects have $\Delta R
\lesssim 0.10$ mag, while those near the completeness limit have $\Delta R
\lesssim 0.20$ mag.  However, a small bias exists, in the sense that
the magnitudes measured on the {\sc NGC 4889} field are about 0.05-mag
brighter at all magnitudes.  If this effect exists for the other
overlap region, it is at a much lower level, and only occurs for the
brighter objects.  This discrepancy is most likely due to flat fielding
errors, and while not completely negligible, these differences are
sufficiently small that we do not attempt to correct for them.  In the
lower panel of Figure \ref{OVERLAPcompare} we plot the color
difference $\Delta (B-R)$ mag versus the total $R$ magnitude, for the
same objects and the same magnitude range of the previous figure. It
is evident that at all magnitudes, the agreement between the color
scales is excellent, with no evidence for bias between fields.

\placefigure{OVERLAPcompare}

\subsection{Combined Photometry Lists}

The calibrated object photometry lists for the four CCD fields
described in Paper I (the three program fields {\sc NGC 4874}, {\sc
NGC 4889}, {\sc NGC 4874 South} and the {\sc Control} field) were
further culled to reduce contamination, according to the following
restrictions.  We discarded objects located within the bright stellar
sequence defined by $r_{-2} < 1.6$ pixels and $R < 19.5$ mag.  (The
pixel scale is $0.53$ arcsec/pixel, thus 1 px = $0.185h^{-1}$ kpc.)
The number of starlike objects in this region are 63, 59, 67 and 64
for the fields as listed above, consistent within Poisson statistics,
as expected.  The fractional contribution of galaxies within this {\em
bright} stellar sequence is negligible (see Paper I). However, for $R
> 19.5$ mag the starlike sequence is also populated with faint compact
galaxies, which are more numerous on the program fields.  To discard
the sequence would impose an extreme bias against these compact
objects.  Thus in this region we depend upon statistical correction
using the {\sc Control} field to account for the contamination by
starlike objects.  Next, the point-like objects with $r_{-2} \lesssim
1.25$ pixels, and all objects with total magnitude below $R = 22.5$
mag (our completeness limit), were culled from the sample.  Then, all
objects which overlapped the physical boundary of the CCD image were
culled from the list, as were objects which overlapped another to such
a degree that photometry was not possible (two to four percent of the
original sample).

Finally, the calibrated object photometry lists for the three cluster
fields were combined to obtain one master list.  Since the three
program fields share a considerable overlapping area, we culled
multiple object detections in the following manner.  For the
North-South overlap region, the final object photometry is taken from
the East edge of the {\sc NGC 4874} field, while for the East-West
overlap region, the object photometry is taken from the North edge of
the {\sc NGC 4874 South} field.  Note that we did not merge the object
lists in the overlap regions, as this would artificially inflate the
detection completeness in these regions.  Two further notes are
relevant.  In the analysis of this paper, we use the observed $(B-R)$
colors, since reddening and extinction are negligible towards the
North Galactic Pole.  As discussed in Paper I, our $2r_1$ aperture
magnitudes are optimized for exponential-profile galaxies, and
therefore they underestimate the total magnitudes for the giant
ellipticals.  The total magnitudes and colors we compute for the 70
brightest cluster galaxies are as accurate as possible given our
analysis method, and we use these values for our analysis in this
paper.  However, other values in the literature may be more accurate,
provided these objects were considered individually using
isophote-fitting techniques.

Over the entire range in $(B-R)$ color and above $R = 22.5$ mag, there
are 3723 objects in the final sample of program field objects, versus
1146 objects on the control field.  For the analysis of Sections 3
through 6, we define a color-restricted subsample which includes
objects with colors in the range $0.7 \le (B-R) \le 1.9$ mag.  These
extremes represent generous limits to the cluster early-type galaxy
(dwarfs and giants) sequence, and reduce contamination from noncluster
galaxies.  This sample includes 2526 program-field objects and 694
{\sc Control} field objects.  The total area of the Coma cluster core
covered by our program fields, minus the area of the overlap regions
and the discarded regions around the cores of the two supergiant
galaxies, is $A_p = 698.44$ arcmin$^2$.  The total area of the {\sc
Control} field is $A_c = 271.13$ arcmin$^2$, thus ${A_p}/{A_c} =
2.5760$.  Then the number density of objects on the {\sc Control}
field (integrated over all magnitudes) is given by $\overline{N_c} =
2.560\pm0.097$ arcmin$^{-2}$ for the color-restricted sample.  Typical
uncertainties at the $R = 22.5$ limiting magnitude are $\sigma(R)
\lesssim \pm0.06$ mag, and $\sigma(B-R) \lesssim \pm$0.12 mag.  
The photometry tables for the full sample of 3723 program field objects
and 1146 control field objects are available electronically from the
first author.

In Section 4, a luminosity function defined by the color-restricted
{\sc Control}-field sample is used to correct the raw luminosity
function observed on the program field.  In addition, we use the
integrated spatial number density to statistically correct the surface
density profiles for our cluster dE galaxies in Section 5.  We do this
with the knowledge that it is {\em only} a first-order correction for
the effect of uniformly distributed foreground stars and noncluster
galaxies.  A single control field provides an accurate estimate (to
within Poisson uncertainties) of the number of genuine Galactic
foreground stars, together with their luminosity and color
distributions.  However, a larger field-to-field variance arises in
the number counts of faint galaxies since they reside predominantly in
clusters and superclusters with apparent angular diameters on the
order of several arcminutes. Thus the field-to-field variance in the
observed number counts will definitely be larger than the Poisson
$\sqrt{N}$ uncertainties.  An average of several randomly offset
control fields would give a more accurate mean number density and
background luminosity function (e.g., Bernstein et al. 1995).  In our
analysis, we instead used an object's {\em color} to discriminate
between cluster and non-cluster galaxies. As illustrated by Biviano et
al. (1995), determination of cluster membership in this manner is
comparable in effectiveness to redshift selection. In this manner we
reduce the field-to-field variance to a level very close to the
irreducible $\sqrt(N)$ amounts.  For a comparison of control-field
number densities with previous studies, refer to Secker \& Harris
(1996).

\subsection{Preliminary Analysis of the Final Sample}

In this section we illustrate the properties of the color-restricted
sample of candidate cluster galaxies defined above, visually, and in
parameter spaces involving the total $R$ magnitude, the $(B-R)$ color,
the intensity-weighted radial moment $r_1$, and the central surface
brightness $I_{\rm c}$.  For purposes of illustration, we show here
grey-scale plots of two small areas (512$\times$512 px$^2$) of the original
$R$-band master CCD images: Figure \ref{Fig3a}(a) is an area northeast
of {\sc NGC 4889}, and Figure \ref{Fig3a}(b) is south of {\sc NGC
4874}.  Here, the numerous dE galaxies are very evident as the
centrally-concentrated low-surface-brightness objects, from moderately
bright objects down to the faintest smudges available to the eye.  To
provide some guidance in separating bright dEs from faint Es, the
brightest few dEs on each image are labeled.

\placefigure{Fig3a}

In Figure \ref{FinalCMD}, we plot color-magnitude diagrams for the
full sample of 3723 program-field objects and for the 1146 {\sc
Control}-field objects.  On the program-field CMD, the dE galaxy
sequence is densely populated and it completely dominates the region
restricted to $0.7 \le (B-R) \le 1.9$ mag.  This galaxy sequence is
completely absent from the {\sc Control}-field CMD, as expected.
In Figure \ref{RADmmnt}, we plot the radial moment $r_1$ (pixels) against
the total $R$ magnitude for the color-restricted sample of 2526 E and
dE galaxy candidates and for the 694 {\sc Control}-field objects.  The
large excess of cluster galaxies, over and above the control field
population, is evident. The {\em most extended dE galaxies} measured
have $5 \lesssim r_1 \lesssim 11$ pixels: for a pixel scale of
$0.185h^{-1}$ kpc/pixel, and with $r_1 = 2r_0$ for exponential profile
objects, this corresponds to an exponential-disk scale radius of
$0.46h^{-1} \lesssim r_{0} \lesssim 1.02h^{-1}$ kpc.  The {\em
faintest dE galaxies which we can resolve} are at $R = 21$ mag: here,
$2 \lesssim r_1 \lesssim 3.25$ pixels, which corresponds to
$0.19h^{-1} \lesssim r_{0} \lesssim 0.30h^{-1}$ kpc.  These values are
consistent with the study of Bernstein et al. (1995), who measure
scale radii in the range $250 \lesssim r_{0} \lesssim 450$ pc for dE
galaxies in the range $22.5 \leq R \leq 23.0$ mag.

\placefigure{FinalCMD}

\placefigure{RADmmnt}

In Figure \ref{SRFbright} we plot the central surface brightness,
$I_{\rm c}$, versus the total $R$ magnitude for the color-restricted
sample of program and control field objects.  The high-$I_{\rm c}$
stellar sequence can be seen to begin at $R = 19.5$ mag. Between $15.5
\le R \le 19.5$ mag, the diffuse sequence of objects is dominated by
cluster dE galaxies.  Below $R \simeq 20$ mag, as we lose our ability
to discriminate between different profiles, the ensemble of objects
merges with the stellar sequence.  Our choice of a limiting magnitude
is evident at $R = 22.5$ mag; while we are complete in {\em magnitude}
above this limit, in {\em surface brightness} we are complete to
$I_{\rm c} \simeq 25$ mag/arcsec$^2$.  Note, however, that we detected
objects with $I_{\rm c} \simeq 26.5$ mag/arcsec$^2$, so as not to bias
our sample against low surface brightness dE galaxies.

\placefigure{SRFbright}

\section{The $R$ versus $(B-R)$ Color-magnitude Distribution}

An analysis of the photometric colors of dwarf (and other) galaxies
can provide valuable information concerning stellar populations of
individual galaxies and populations of galaxies (Caldwell \& Bothun
1987; Evans, Davies \& Phillipps, 1990; Garilli et al. 1992; Cellone,
Forte \& Geisler 1994).  In Secker and Harris (1996), the $(B-R)$
colors were used primarily as a discriminator to eliminate
contamination due to non-cluster galaxies and foreground stars.
However, the $(B-R)$ color index, like $(B-I)$ and the Washington
$(C-T_1)$ color, is also a sensitive and accurate estimator of the
total heavy element abundance (metallicity) for old stellar
populations (Geisler \& Forte 1990; Couture, Harris \& Allwright 1991,
1992; Held \& Mould 1994; Secker 1996), and an analysis of the CMD is
warranted.  Note that while both age and metallicity affect absorption
features in the spectra (and therefore the integrated color) of
stellar populations, the effect of metal abundance dominates over age
effects in the observed $(B-R)$ color of old populations such as dE
galaxies (Worthey 1994).  Thus in our sample of galaxies, we assume
that the redder galaxies are, on a relative scale, more metal rich
than the bluer galaxies (Secker 1996).

Figure \ref{FinalCMD} illustrates the color restriction (i.e., $0.7 \leq (B-R) 
\leq 1.9$ mag) which represents generous limits on the early-type galaxy 
sequence: the most probable members of the Coma cluster are included,
while the large contribution from red background galaxies is excluded.
It is this data sample which we analyze here, and in the right panel
of Figure \ref{CMDanalysis}, we reproduce this subset of the total
cluster field CMD.  The dE galaxy sequence begins at $R \simeq 15.5$
mag, with a mean color of $(B-R) \simeq 1.54$ mag.  This galaxy
sequence shows a strong trend for fainter dE galaxies to have (on
average) a bluer color.  This trend was also observed for the brighter
dEs in Coma by Thompson \& Gregory (1993) and Biviano et
al. (1995). To quantify this trend, we calculate the trimmed
median $(B-R)$ color (the solid circles) in one-magnitude bins over
the entire luminosity range.  The upper portion of the galaxy sequence
is well defined and has very little apparent contamination.
Considering only this bright part of the galaxy sequence (i.e.,
$14.0 < R < 18.5$ mag), we perform a least-squares fit of a straight
line to the median color values.  The resulting best-fit regression
line is plotted (solid line) over the valid range in the left panel of
Figure \ref{CMDanalysis}, and it clearly provides an excellent fit to
the data.  This regression line is given by

\begin{equation}
(B-R) = (-0.056\pm0.002) R + (2.41\pm0.04).
\end{equation}

Also plotted in the left panel of Figure \ref{CMDanalysis} are open
circles, which correspond to the mean color of control-field objects,
calculated in the same 1 magnitude bins (but only below $R = 18.5$
mag).  The dashed line is an extension of the upper line; it is not a
fit to any of the fainter data points.  The control-field sample is
dominated by faint noncluster galaxies, and this is apparent here; the
mean colors given by the open circles are in all cases redder then the
extrapolated dE galaxy sequence (dashed line).  At magnitudes fainter
than $R = 18.5$ mag, the dE sequence spreads due to photometric error
and it merges with the multitude of noncluster galaxies, which are
clearly the dominant population (in number fraction) at these
magnitudes.  The effect is that the program-field mean color values
(solid circles) deviate from the dashed line, redward, towards the
control-field mean color values.  Even using a trimmed median which is
by definition less sensitive to outliers cannot yield the true mean
colors for the faint end of the sparse dE galaxy sequence, as it is
the dEs which are the outliers of a color distribution function
dominated in number by non-cluster galaxies.  Below, the question we
address is whether this blue trend observed for the brighter dE
galaxies is continuous and linear over the full magnitude range, down
to our limit at $R = 22.5$ mag.  We determine that it is, and that the
apparent flattening observed is caused by the redder noncluster
galaxies which compose the majority of our program field sample at
these faint magnitudes.

\placefigure{CMDanalysis}

We adopt a method which is similar to Secker et al. (1995) and Kuhn et
al.  (1996), to derive a decontaminated version of the program-field
CMD.  Our procedure first creates binned versions of the program-field
and the control fields CMDs, preserving the two-dimensional nature of
the data, something which is impossible using a scaled and subtracted
histogram.  Once the CMDs are binned, the number counts in the
control-field CMD are scaled by the ratio of areas, $A_p/A_c = 2.5760$
(Section 2.3), and subtracted from the program-field CMD.  In this
manner we correct for a localized excess of control-field objects in
the CMD of the program fields.  We define $P(c,m)$ and $C(c,m)$ to
represent the number of objects in the program-field and control-field
CMDs respectively, computed in bins of height 1.0-mag in total $R$
magnitude and width of 0.1 magnitudes in $(B-R)$ color.  Then the
number of objects in each bin of the control-field subtracted CMD
$N(c,m)$ is given by

\begin{equation}
N(c,m) = P(c,m) - \frac{A_p}{A_c}C(c,m),
\end{equation}

\noindent where $N(c,m)$ is computed over the range $12.5 \ge R \ge 22.5$ 
mag and in the color range $0.7 \ge (B-R) \ge 1.9$ mag.    The 
uncertainty on $N(c,m)$ is given by the $\sqrt{N_{ij}}$ counting 
uncertainties for the program- and control-field number counts added 
in quadrature.

In the right panel of Figure \ref{CMDanalysis}, we plot the binned,
scaled and subtracted CMD, $N(c,m)$, corresponding to what we expect
for Coma cluster members.  There is one vertical line segment for each
$N(c,m)$ bin, with a height proportional to the number of objects in
that bin.  In this case, the solid circles represent the mean
corrected color, calculated for each 1-mag bin by
averaging over all color bins, weighted by the number of objects in
each bin.  The associated errorbars are not intended to convey
accurate uncertainty estimates; they simply represent the spread of
objects about the median color (i.e., the semi-interquartile range) in
the initial program-field CMD.  Reproduced in the right panel is the
solid regression line form the left panel, together with the
extrapolated dotted line.  That is, we did not perform a separate
least-squares fit to the new mean color values.  The important result
here is that {\em this regression line provides an excellent fit to
the corrected mean color values for the dE galaxy sequence, over the
full range $15.5 \le R \le 22.5$ mag}.  Any apparent reddening in the
left panel can therefore be attributed entirely to contamination of
the cluster sample by noncluster galaxies and foreground stars.  Note
that redward of the dE galaxy sequence and fainter than $R = 19.0$
mag, the number of cluster galaxies is consistent with zero, though
there is considerable scatter, including a number of negative values
for $N(c,m)$.  This occurs because the control-field CMD provides only
an estimate for the number and distribution of objects in color and
magnitude expected on the program field, and it will not (and should
not) subtract out perfectly.

The color-magnitude trend we observe in the Coma dEs is quite
certainly a continuation of the sequence defined by the early-type
cluster giants, and it is very similar to that observed recently in
Fornax (but see Evans, Davies \& Phillipps 1990 for a different
result).  Cellone, Forte \& Geisler (1994) plot a $T_1$, $(C-T_1)$ CMD
(Washington filters) for a sample of 14 Fornax dEs and notice this
color-magnitude correlation.  From a linear fit to their tabulated
data points, we estimate a decrease of $0.078\pm0.029$ mag in
$(C-T_1)$ for each magnitude decrease in total $T_1$ magnitude.  As
stated earlier, we assume that integrated $(B-R)$ colors measured for
our sample of dEs reflect metallicities of the underlying stellar
populations, in the sense that redder dEs are more metal rich than
bluer dEs.  This is supported by Cellone, Forte \& Geisler (1994), who
find that the metallicity-insensitive color index $(M-T_1)$ is
essentially constant with magnitude (within the scatter) for the same
sample of 14 dEs in Fornax.  For Fornax Es, Caldwell \& Bothun (1987)
find an decrease of $0.08\pm0.012$ mag in $(U-V)$ for each magnitude
decrease in total $B$ magnitude.  This color-magnitude trend may in
fact extend fainter to their sample of 11 dEs, but with a much larger
scatter.  In order to compare the color-magnitude relationship for our
Coma cluster dEs with those in Fornax, we adopt color-metallicity
relationships derived from metallicities and integrated colors for
Galactic globular clusters (from Harris 1996).  These are given by
[Fe/H] $= (3.44\pm0.09)(B-R)_0-(5.35\pm0.10)$ and [Fe/H] $=
(1.92\pm0.05)(U-V)_0-(2.97\pm0.04)$, and are valid over the range
$-2.4 \lesssim$ [Fe/H] $\lesssim -0.2$.  From Geisler \& Forte (1990),
we adopt the color-metallicity relationship [Fe/H] $= 2.35(C-T_1)_0 -
4.39$, which is also calibrated against Galactic globular clusters
over a similar metallicity range.  For our sample of Coma dEs, we
therefore obtain a metallicity change with magnitude of
$\Delta$[Fe/H]$/ \Delta R = -0.19\pm0.01$ dex.  For the Fornax dEs, we
calculate $\Delta$[Fe/H]$/ \Delta T_1 = -0.18\pm0.07$ dex for the
Cellone et al. (1994) sample, and estimate $\Delta$[Fe/H]$/ \Delta B =
-0.16\pm0.07$ dex for the Caldwell \& Bothun (1987) sample.  These
slopes are formally consistent, suggesting that a very similar
phenomenon has been at work in both these cluster environments.

These color-luminosity correlations have a simple interpretation
within the context of dE galaxy formation.  As described by Dekel \&
Silk (1986), the brighter (and more massive) dE galaxies have deeper
gravitational potential wells, and therefore better able to retain the
interstellar gas which became super heated and metal enriched during
initial stages of the star-formation epoch.  Although star formation
in early-type galaxies is believed to be short lived (i.e., $\lesssim
10^9$ years), this is sufficient time for several generations of
massive stars to form out of this metal-enriched gas, leading to an
observed stellar population with is enhanced in metals (redder)
compared to the fainter dE galaxies.

In Figure \ref{CMDanalysis}, the spread in the dE sequence, about the
fiducial line and at any magnitude, exceeds that which can be
attributed to photometric error.  For the restricted (bright-end)
range $15.5 \leq R \leq 18.0$ mag, the total width is nearly constant
at $\sigma_{\rm obs} = 0.07$ mag.  And from Paper 1, the typical
photometric error in the color estimate for our sample of objects is
$\sigma_{\rm err} \lesssim 0.025$ mag.  We conclude that the intrinsic
width in the color of dE galaxies (at bright magnitudes) is
$\sigma_{\rm int} = (\sigma_{\rm obs}^2 - \sigma_{\rm err}^2)^{1/2}
\simeq 0.065$ mag.  This intrinsic scatter in the dE
galaxy metallicity could reflect fragment-to-fragment differences in
an early epoch of pre-enrichment, or local variations in the density
of the intergalactic gas pressure (Babul \& Rees 1992; Secker 1996).

The diffuse intergalactic component of the Coma cluster light is
estimated to be on the order of 25 -- 30 percent of the total light
(Thuan \& Kormendy 1977; Melnick, White \& Hoessel 1977; see also Uson
et al. 1991 for a more recent discussion of the difficulties inherent
in measuring this intracluster light).  Thuan \& Kormendy (1977)
report that for Coma, the diffuse light appears to be bluer than the
light of the giant galaxies, at least within 14 arcminutes of the
core. This is supported by Mattila (1977), whose photoelectric
photometer measurements yield $(B-V) = 0.54\pm0.18$ mag, in an area
free of galaxies. Though quite blue, the Mattila (1977) measurement is
consistent with the most metal-poor old stellar populations such as
globular clusters, which are in the range range $0.55 \lesssim (B-V)
\lesssim 0.85$ mag,

One possible origin of this low surface brightness diffuse light is
stars tidally stripped from the outer regions of their parent
galaxies.  As illustrated in Figure \ref{CMDanalysis}, the faint dE
galaxies are the most metal poor (bluest) galaxies in this
environment, and as discussed in Section 5, they are also the most
easily disrupted, and we have evidence that a significant population
of these dE galaxies have been tidally destroyed in the cluster core.
Therefore, it is interesting to compare the color of this intracluster
light with the median color for faint dE galaxies.  From Figure
\ref{CMDanalysis}, an approximate median color for the dE galaxies (at
$R = 22.5$ mag) is $(B-R) \simeq 1.15$ mag.  To convert this to an
equivalent $(B-V)$ color, we use the $(B-V)-(B-R)$ color-color
relation for individual halo giants of Durrell, Harris \& Pritchet
(1994).  If we assume this is approximately correct for the integrated
colors of dE galaxies, our $(B-R)$ dE color corresponds to $(B-V)
\simeq 0.70$ mag, which is within the color range found by Mattila
(1977) and similar to an intermediate-metallicity globular cluster
([Fe/H] $\simeq -1$).  Thus, our CMD analysis is consistent with a
major component of this diffuse intergalactic light originating as
stars tidally stripped from numerous faint dE galaxies.  If so, the
mean color of this diffuse light should be consistent with that of the
faint dE galaxies at all clustercentric radii.  A more accurate
measurement of the diffuse light would be of great interest.

\section{The Galaxy Luminosity Function}

The probability distribution function of luminosity for galaxies
in a cluster is a fundamental observable quantity
which places physical constraints on galaxy formation and subsequent
dynamical evolution. From a CDM, self-similar, stochastic
model for galaxy formation (Press \& Schechter 1974), Schechter (1976)
derived a robust analytical form to parameterize the luminosity
function (LF),  

\begin{equation}
\Phi(L)dL = \phi_{\star} \left( \frac{L}{L_{\star}} \right)^{\alpha} \exp 
\left( -\frac{L}{L_{\star}} \right) \frac{dL}{L_{\star}}.
\end{equation}

Here $L_{\star}$ is a characteristic luminosity at which the slope of
the Schechter function changes fairly abruptly, $\alpha$ is the slope
of the power law for $L \ll L_{\star}$, and $\phi_{\star}$ provides
the normalization (number per unit volume).  For a given ensemble of
galaxies, the Schechter LF provides a convenient model for the
comparison of galaxy populations in differing environments.  Recent
observations strongly dictate that there is no universal LF: not only
does it differ between field and cluster environments, there is
significant observed variation between different cluster environments.
While it appears that individual galaxy morphological classes have
unique LFs, their relative contribution to the composite LF is a
product of environment (Binggeli, Sandage \& Tammann, 1988).  Although
the Schechter function does not adequately describe the composite
galaxy LF (giants plus dwarfs) for the Coma cluster, it does provide a
good fit to the dE luminosity function (Sandage, Binggeli \& Tammann
1985; Thompson \& Gregory 1993; Biviano et al. 1995; Secker \& Harris
1996).

The deepest study of the LF for the Coma cluster core
is that by Bernstein et al. (1995).  Over the range in magnitude of
$15.5 \leq R \leq 23.5$ mag, they determine that the slope of the
faint-end power law is $\alpha = -1.42\pm0.05$, with a constant
faint-end slope over this entire magnitude range.  As well, Thompson
\& Gregory (1993) find that $\alpha \simeq -1.43$ for the faint-end of
the LF in Coma.  These values are consistent with findings for both
the Virgo and Fornax clusters (Binggeli, Sandage \& Tammann 1985;
Sandage, Binggeli \& Tammann 1985; Ferguson \& Sandage 1988).
However, the LFs measured in the cores of other rich galaxy clusters
(not including Coma) have been found to be much steeper at their
limiting magnitudes: De Propris et al. (1995) find $\alpha \simeq -2$
for A2052, A2107, A2199 and A2666 ($z = 0.035, 0.042, 0.031, 0.026$),
Driver et al. (1994) find $\alpha \simeq -1.8$ for A963 ($z = 0.21$), 
and Wilson et al. (1997) find $\alpha \simeq
-2$ for A665 and A1689 ($z = 0.18$).  Here, we determine the slope for the
faint-end of the LF {\em as defined by our sample of dE galaxies in
the cluster core}, and compare this to the findings cited above.

We consider the color-restricted subsample of Section 2, which consists 
of 2526 program-field objects (some fraction of which are E
and dE galaxies in the cluster), and 694 control-field objects;
objects in both samples satisfying $R \leq 22.5$ mag and $0.7 \leq
(B-R) \leq 1.9$ mag.  As is commonly done, we plot the galaxy LF as
the number of galaxies per unit {\em magnitude}.  A change of
variables involving the definition of magnitude, its derivative and an
expression for the luminosity, yields

\begin{equation}
\Phi(m) = \kappa \left\{ 10^{0.4(m_{\star}-m)} \right\}^{(\alpha+1)}
\left[ \exp \left( -10^{0.4(m_{\star}-m)} \right)  \right],
\label{eq:magsch}
\end{equation}

\noindent where $\kappa = \phi_{\star} \ln 10 / 2.5$.  In
Table \ref{tbl2} we summarize the relevant LFs (in 0.5-mag bins): the
raw counts for the Coma program fields, the area-scaled counts for the
control field, and the final control-field subtracted
completeness-corrected galaxy LF.  The completeness function we used
is that given by (\ref{eq:compfcn}) in Section 2.2.  The error
estimate for the final galaxy LF is derived from the Poisson
uncertainties for both the control-field and program-field counts,
added in quadrature.

In Figure \ref{FaintendLF} we plot this LF over over the magnitude
range $15.5 \leq R \leq 22.5$ mag, the range of interest for the
faint-end slope analysis (i.e., $R \simeq 15.5$ mag marks
approximately the onset of dE galaxies).  In the faintest three bins
(i.e., for $R \gtrsim 21$ mag) the dwarf galaxy LF appears to level
off, an effect which cannot be attributed to magnitude incompleteness.
These three bins are, however, relatively uncertain, as they result
from a relatively large background component.  By themselves,
therefore, they do not provide compelling evidence for a leveling-off
in the dE galaxy LF.  However, Lobo et al. (1997) find evidence for a
variation in the faint-end slope of the galaxy LF (steeper away from
NGC 4874 and NGC 4889), and in Section 5, we describe further evidence
for a paucity of faint dE galaxies in the cluster core, beginning near
$R = 19$ mag. Thus these features of the LF may be a manifestation of
tidal disruption (or inhibited formation) of faint dEs in the dense
cluster core.

The faint-end slope of our galaxy LF was obtained via weighted
least-squares regression over the magnitude range $15.5 \leq R \leq
22.5$ mag.  The slope of the regression line is given by $\Delta \log
(N)/ \Delta m = 0.166\pm0.019$; this best-fit line and its
uncertainties are plotted over the corresponding magnitude range on
Figure \ref{FaintendLF}.  This slope corresponds to $\alpha =
-1.41\pm0.05$, and it is this value which we adopt here for the
faint-end slope. (This is the value that was used in Secker \& Harris
1996 for their modeling and analysis of the composite galaxy LF.)  We
also analyzed a magnitude-restricted subset of our galaxy LF (the
region over which our LF rises with the steepest slope, $17 \le R \le
21.0$ mag), for which the resulting least-squares regression line
(hatched line in Figure \ref{FaintendLF}) yields $\alpha =
-1.49\pm0.07$.  Our adopted value for the faint-end slope of the Coma
core LF is entirely consistent with the slopes obtained by Thompson \&
Gregory (1993) and Bernstein et al.  (1995), and it is consistent with
the average of the core faint-end slopes ($\alpha = -1.55\pm0.12$ for
the NGC 4874 and NGC 4889 regions) measured by Lobo et al. (1997).
However, these measured faint-end slopes for Coma all differ
significantly from the much steeper values measured for other clusters
by De Propris et al. (1995), Driver et al. (1994) and Wilson et
al. (1997), even though all of these LF measurements were made in the
cluster cores, where we may expect the shallowest faint-end slope
(Lobo et al. 1997).  In addition, the seven clusters studied by De
Propris et al. (1995), Driver et al. (1994) and Wilson et al. (1997)
vary in Bautz-Morgan class (I - III), span a full range in richness
(richness-class 0 up to 5), and all but one (A1689) has a dominant cD
galaxy.  Further deep imaging studies of rich galaxy clusters are
essential to map all of the cluster parameter space and pinpoint the
origin of these steep cluster luminosity functions.

In Figure \ref{ModelLF}, we plot the net Coma galaxy LF from Table
\ref{tbl2} (solid circles) over the complete range $12 \le R \le 22.5$ 
mag.  We refer to this as the composite LF, since it includes
contributions from both the cluster giants and dwarfs.  The
contribution of the giants ($R \lesssim 16$ mag) to this composite LF
is immediately obvious (within the scatter of small-number statistics)
as a log-normal distribution peaked near $R = 14.5$ mag. (Note that
the total $R$-band magnitudes we derive for cluster giants may be
underestimated; see Section 2.3 and Paper I.)  In Secker \& Harris
(1996) we modeled this composite LF as a sum of a Gaussian function
for the giants and a Schechter function for the dEs (cf. Thompson \&
Gregory 1993; Biviano et al. 1995).  A least-squares fit of this
Gaussian plus Schechter model for several values of the Gaussian
dispersion (in each case with the faint-end slope fixed at $\alpha =
-1.41\pm0.05$) yielded values for the characteristic magnitude
$R^{\star}$, the peak of the Gaussian distribution, and for the
relative normalization factors; refer to Secker \& Harris (1996) for a
full table of model parameters.  The range of Gaussian dispersions
which we considered to model the giant galaxies is $\sigma = 0.8-1.2$
mag, constrained by the literature.  In Figure \ref{ModelLF}, the
solid line illustrates the best model fit ($\sigma = 0.8$ mag and
$\alpha = -1.41$), and the dotted line illustrates the corresponding
decomposition of this model fit into the two separate components.
These are formally the best-fit parameters, although models with
higher $\sigma$ ($1.0 - 1.2$ mag) also provide adequate fits.

\placefigure{ModelLF}

\section{Galaxy Spatial Distributions}

The Coma cluster core is dominated in luminosity by two supergiants,
the elliptical NGC 4889 ($B_{T} = 12.53\pm0.11$, $(B-V)=1.04\pm0.01$)
and the cD NGC 4874 ($B_{T} = 12.63\pm0.11$, $(B-V)=0.95\pm0.02$) (de
Vaucouleurs et al. 1991).  Like many Abell clusters, Coma has
significant substructure (Fitchett
\& Webster 1987; Escalera, Slezak
\& Mazure 1992; Davis \& Mushotzky 1993; Mohr et al. 1993; White, 
Briel \& Henry 1993; Colless \& Dunn 1996). In terms of optical
morphology, Coma is a Rood \& Sastry (1971) B-type (binary) with NGC
4889 and NGC 4874 both obvious centers of galaxy concentration in the
core.  However, NGC 4874 appears special for two reasons.  First, it
is located at the peak of the diffuse X-ray emission in the cluster
core (White, Briel \& Henry 1993) and it has a strong nuclear radio
source characteristic of many central giant ellipticals (Harris 1987).
Second, while both NGC 4874 and NGC 4889 have an associated globular
cluster system (GCS), only NGC 4874 has the high (M87-like)
specific-frequency GCS which to date has been found only in cD-like,
centrally-dominant ellipticals (Harris 1987; Harris, Pritchet
\& McClure 1995).  For these reasons we adopt the position of NGC 4874
as the cluster center.

The distribution of giant galaxies is well described by the {\em
morphology-density} (or the {\em morphology-clustercenter radius})
relationship: the morphological fraction of galaxies at any location
in a galaxy cluster is independent of the cluster's global parameters
(e.g., richness), and is governed primarily by the local projected galaxy
density (Dressler 1980) (or the clustercentric radius; Whitmore,
Gilmore \& Jones 1993).  The morphology-density relationship also
holds true for dwarf galaxies, for both the Virgo cluster (Binggeli,
Tammann \& Sandage 1987; Binggeli, Tarenghi \& Sandage 1990; Ferguson
\& Binggeli 1994) and the Coma cluster (Thompson \& Gregory 1993).
These studies found that the early-type dE galaxies compose the
largest number fraction of galaxies in the dense environment of the
cluster core.  For Coma dEs, Thompson \& Gregory (1993) determined
that brighter dEs follow the distribution of the early-type giant
galaxies throughout the cluster, rising steeply towards the center.
They found that while the faint dEs are distributed in the same manner
as the bright dEs in the outer regions, a deficit of faint dE galaxies
was detectable within 0.35 arcmin of cluster center.  Our goal is to
use our large sample of program-field objects and our estimate of the
background number density of galaxies to analyze the dE galaxy radial
number-density distribution, and compare this to the radial profile
for the early-type giants sharing the cluster core.

In Figure \ref{SpatialE}(a) we plot positions and magnitudes for 280
{\em bright} cluster galaxies, a subsample selected to have $0.7 \leq (B-R)
\leq 1.9$ mag, and $R \leq 19.0$ mag (giants plus bright dEs).
Immediately noticeable is the strong clustering of bright galaxies
around NGC 4874. In Figure \ref{SpatialE}(b) we plot the
subsample of 2246 {\em faint} objects selected by color to $0.7 \leq (B-R)
\leq 1.9$ mag, and limited in magnitude to $19.0 < R \ge 22.5$
mag. While this sample includes the faint dE galaxies, it is dominated
in numbers by the spatially-uniform background population.  In
contrast to the giants and bright dEs in (a), there is no immediately
apparent fall off in the density of faint dEs, the first indication of
a significantly larger core radius.  Note that our sample is
incomplete within a radius of 150 pixels around both NGC 4874 and NGC
4889, and these regions are excluded in the following analysis.

\placefigure{SpatialE}

The radial number density profiles for the two samples of galaxies
were derived by (i) summing the completeness-corrected number counts
within radial annuli, (ii) normalizing these counts to the usable
area of the annuli, and (iii) correcting for the number density of
background objects calculated in the relevant color and magnitude
range on the {\sc Control} field. In Figure \ref{SpatialE} the
concentric circles (dotted lines) designate the boundaries of the
radial annuli used to compute the radial profile for both galaxy
samples.  The inner annulus begins 1.325 arcmin from NGC 4874, the
outer annulus ends at 23.325 arcmin, and the region in between is
divided evenly into 2 arcmin bins (except for the outer annulus; see
Table \ref{tbl3}). These are the same radial bins used for the
analysis of the radial color distribution in Secker (1996).
Concerning the corrections for spatial incompleteness, the actual area
of the annuli used in calculations is that which intersects useable
regions of the CCD images, as designated by the solid line boundary in
Figure \ref{SpatialE}. Also plotted in Figure \ref{SpatialE} (a) and
(b) are incomplete regions due to NGC 4787, NGC 4889, and a very
bright star north of NGC 4874.  In Figure \ref{SpatialE} (b) the small
open circles are located at the positions of extended giant galaxies,
and represent a first-order correction to the resultant spatial
incompleteness.  All incomplete regions are taken into account when
computing the area of the annuli.

\placetable{tbl3}

The spatially varying magnitude completeness corrections required
additional simulations.  The completeness function, $f(m)$, of Section
2 is a global average, derived from a sample of objects over a large
and variable region of the cluster core.  It was appropriate to use
$f(m)$ to correct the galaxy luminosity function, since the galaxy LF
represents a sum of all objects detected over these same variable
regions (Section 4).  However, near NGC 4874, the radial number
density profile is affected by a changing object detection level:
faint non-stellar objects are more likely to be missed due to the
higher Poisson noise.  To quantify this effect we performed
artificial-star tests, and derived a separate magnitude completeness
function for each of the four inner radial annuli. For the outer 6
annuli, we adopted the completeness function calculated for the fourth
annulus. (At these larger radial distances, light gradients from NGC
4874 no longer affect the detection completeness, which is
predominately due to sky noise.)  For these simulations we used
stellar profile objects, since at these faint magnitudes, the majority
of the objects (stars and faint galaxies alike) have profiles close to
stellar.  We added a total of 1800 stars distributed as a power-law
over the magnitude range $16 < R < 23$ mag, and used a uniform spatial
distribution.  These stars were added to the {\sc NGC 4874} field
$R$-band master image in six batches of 300 stars each; in total, 209,
388, 434 and 719 stars were added into the innermost, second, third
and fourth annuli.

\placefigure{Artstarcomp}

The resulting completeness functions for the four annuli, $f(R)$, 
are illustrated in Figure \ref{Artstarcomp}.  Here, the
solid circles represent the completeness fraction, and the associated
uncertainties were derived using Bolte's (1989) formula, given in
Section 2 as equation (\ref{eq:comperr}).  The function $f(R)$ was
computed using the number input in each 0.5-mag bin, and the number of
these objects detected at any magnitude. (That is, we do not care
about the effects of bin jumping, as we are only concerned with detection
completeness.)  For the innermost annulus, the detection completeness
is $f(R) = 0.80$ at our $R=22.5$ mag limit.  The other three annuli
are only slightly different.  To these solid circles we chose
to fit an analytic function, to smooth out the effects of small-number
statistics, and to provide a convenient interpolation between the
data points.  We adopted a two-parameter function due to C.J. Pritchet,
for which $\alpha$ describes the shape of the detection completeness
curve, and for which $m_{\ell}$ is an effective (i.e., 50 percent
complete) limiting magnitude:  

\begin{equation}
f(R) = \frac{1}{2} \left[1-\frac{\alpha(R-m_{\ell})}
{\sqrt{1+\alpha^2(R-m_{\ell})^2}} \right].
\end{equation}

\noindent With the parameters illustrated in the four panels of Figure
\ref{Artstarcomp}, it provides a good fit (dotted line) to these completeness
functions.  We then used the detection completeness functions
derived from these artificial star tests to correct the number counts
in each annulus in each magnitude bin.

\placefigure{RDFfinal}

Finally, the completeness-corrected radial number density
distributions are illustrated in Figure \ref{RDFfinal}, in a plot of
$\log(N/{\rm arcmin}^2)$ versus $\log(R_{\rm cc}/{\rm arcmin})$.  The
number densities are plotted at the geometric mean radius (i.e., $r =
\sqrt{r_i r_j}$) of the annulus, and are corrected for the mean 
{\sc Control}-field number density of objects.  Within the color range
$0.7 \le (B-R) \le 1.9$ mag and for $R \le 19.0$ mag, the
control-field number density is $0.070\pm0.016$ objects per
arcmin$^2$.  Within the same color range, and for $19.0 < R \le 22.5$
mag, the control-field number density is $2.490\pm0.096$ objects per
arcmin$^2$.  It is clear in the upper panel of Figure \ref{RDFfinal}
that the number density profile of the early-type giants and bright
dEs continues to rise into our innermost bin at $R_{\rm cc} = 2.1$
arcmin, while the density profile for the faint dE galaxies (lower
panel) is very flat throughout this radial range.  It is the relative
differences of the core radii for these two samples of galaxies which
we are primarily interested in, and below we quantify the shapes of
these radial profiles using model fits to the data. Note first,
however, that the outermost three points in both radial profiles
correspond to only a small fraction of the total cluster area at that
radius, and are thus very spatially incomplete around the annulus.

To both of these projected radial surface density profiles we fit an
isotropic, single mass King model (King 1966).  While these King
models were originally proposed to model globular star clusters, they
have also been used successfully to model galaxy clusters (Bahcall
1977; Binggelli et al. 1987).  Our main interest in these models is
that they fit well to our radial density profiles, and thus provide a
convenient method to compare the core radius and central surface
densities between our two galaxy samples.  In practice, we numerically
compute King models for various values of the concentration index,
which we then shift vertically and horizontally until they best fit
the observed data (McLaughlin et al. 1995). The model which, when
shifted, provides the minimum $\chi^2$ statistic, is taken to be the
best fit to the data.  Note that the concentration index $c$ is based
upon the tidal radius $r_t$, the estimation of which requires a large
extrapolation to our radially-limited data set.  Thus the values for
$c$ quoted below are less certain than the estimated central density
and core radius.

For the bright galaxy sample, the best-fit King model (with
$\chi^2_{\nu} = 1.43$) has a central surface density $\Sigma_0 =
0.53$ arcmin$^{-2}$, a concentration index of $c = 0.28$, and a core
radius $R_{\rm c} = 13.71$ arcmin, corresponding to $R_{\rm c} =
287.1h^{-1}$ kpc.  Our measured core radius is larger than the value
of 9 arcmin obtained by Kent \& Gunn (1982), with the difference most
probably arising from our limited radial coverage.  For the faint dE
galaxies, the best-fit King model (with $\chi^2_{\nu} = 1.37$) yields
a central surface density $\Sigma_0 = 1.17$ arcmin$^{-2}$, a
concentration index of $c = 0.28$, and a core radius $R_{\rm c} =
22.15$ arcmin, corresponding to $R_{\rm c} = 463.8h^{-1}$ kpc.  These
best-fit King models are plotted along with the data points in Figure
\ref{RDFfinal}.  There is a significant difference between these fits,
in the sense that a least-squares fit of the bright-sample model to
the faint-sample data (or vice versa) yields an extremely large value
for the reduced chi-squared parameter.

As mentioned above, Thompson \& Gregory (1993) tentatively proposed
that the core of the Coma cluster is deficient in the number of faint
dE galaxies, and they argued that dynamical effects (tidal disruption)
in the rich environment of the cluster core could be responsible for
partially destroying this population of faint dE galaxies. However,
they cautioned that this relatively flat distribution for the faint dE
galaxies might simply be a manifestation of a uniformly distributed
population of faint noncluster galaxies.  Our study indicates that
this flat inner distribution is genuine.  Scaling the radial profile
of the bright sample of galaxies to that of the faint dE population
only, we see that the number density of the bright galaxies continues
to rise farther into the core of the cluster (i.e., it has a smaller
core radius), so in this sense the cluster core is indeed deficient in the
numbers of faint dE galaxies.  This is consistent with the results
of Lobo et al. (1997), who determine that the faint-end slope of the 
galaxy luminosity function is shallower in the Coma cluster core,
when compared to the cluster as a whole.

A further note is warranted here with respect to the destruction of
dEs in the cluster core. Moore, Lake \& Katz (1997) (following from
Moore et al. 1996) describe the formation of dE galaxies in dense
clusters as morphological transformations due to gravitational
interactions (``galaxy harassment'').  They suggest that the radial
color gradient detected in our sample of bright Coma cluster dE
galaxies (Secker 1996) is a result of the destruction of faint blue
dEs in the core.  They believe that the absence of these faint blue
dEs skews the mean galaxy color, resulting in a mean color which is
redder towards the cluster center.  We think that their argument is
not likely to be correct for the following reasons: (a) we find that
the sample of bright dE galaxies ($R \le 19.0$ mag) has the {\em same
radial distribution as the giant galaxies}, suggesting that no
significant tidal disruption has occurred for them. It is only the
faint dE galaxies (i.e., with $19 < R \le 22.5$ mag) which have a
significantly different radial number density distribution, a probable
consequence of tidal disruption.  (b) The color gradient noted by
Secker (1996) exists over the full radial range, and it is not limited
to the inner 100 kpc, as suggested by Moore et al. (1997).  Thus,
while we agree that the paucity of faint, fragile, blue dEs would
produce a radial color gradient in the inner 100 kpc, the color
gradient measured by Secker (1996) for the bright dEs can not be
attributed solely to tidal disruption, and it likely represents a
variation in the mean dE metallicity with clustercentric radius.

One final comment concerns the effect of the different spatial
distributions between luminosity-selected samples on the computation
of the early-type dwarf-to-giant ratio (EDGR).  Secker \& Harris
(1996) used the color-restricted sample defined in Section 2.3 to
estimate the EDGR for the Coma cluster.  To a limiting magnitude of $R
= 18.6$ mag, they derived a value of $1.80 \pm 0.58$, consistent with
a value computed by Thompson \& Gregory (1993), and consistent with
values for the Virgo cluster.  The EDGR computed by Secker \& Harris
(1996) is based on observations of the cluster core only, and one may
question whether this is representative of the Coma cluster as a
whole.  Ferguson \& Sandage (1991) find no evidence to support a
significant difference in their computed EDGR with a variation in
clustercentric radius.  They determined this by computing EDGR values
in annuli of width one degree as a function of radius outwards from
the Virgo cluster center. As such, they conclude that only a single
value for the EDGR is necessary to characterize the Virgo cluster, and
postulate that this is true for all individual galaxy clusters.  To
first order, this is true in the Coma cluster as well, but given the
difference in spatial distributions between the cluster giants and the
faint dEs determined above, this point is worth further consideration.

For large radial distance from the cluster center, Thompson \& Gregory
(1993) find that at about 20 arcmin from the center (NGC 4874), the
powerlaw falloff of the faint dE galaxies is consistent with that of
the bright dEs and the E+S0 galaxies.  Since the number of faint dE
galaxies decreases in proportion to the number of bright galaxies, the
EDGR computed at any limiting magnitude should be constant at any
radius.  However, this is not true for the innermost regions of the
core, where the number density of bright dE and E+S0 galaxies
increases more rapidly than does the number density of faint dE
galaxies.  Thus a calculation for the EDGR limited to the central core
of the Coma cluster, as our value is, may underestimate the EDGR for
the galaxy cluster as a whole. However, this underestimation will only
occur if the dE galaxy limiting magnitude (to which the Schechter
function is numericallythe integrated) includes the faintest dE
galaxies.  In Secker \& Harris (1996), the first two values for the
EDGR in the Coma cluster core, i.e., to limits of $R = 18.6$ and 19.6
mag, are therefore representative of the Coma cluster as a
whole. However, their limits of $R = 20.6$ and 21.6 mag are
sufficiently faint that they include the more smoothly distributed
faint dE galaxies, and these estimates of the EDGR value may not be
applicable to the cluster as a whole.

\section{Projected (Bound) Companions to Early-type Giants}

In the previous section we analyzed the radial number density
distribution of dE galaxies within the cluster core, which depended
only on number counts in radial annuli and magnitude bins.  In this
section we use the dE galaxy locations to estimate the number of dwarf
galaxies which are {\em gravitationally bound companions} of
early-type giants. To do this we use number counts of dE candidates
projected near a cluster giant and subtract the {\em local} mean dE
number density.  Dwarf elliptical galaxies which formed along with the
parent galaxy should be useful tracers of the extent of the giant's
dark halo, as both the dEs and dark halos are affected by tidal
stripping (Vader \& Chaboyer 1992; Binggeli 1993).  Within the Coma
cluster, there are at least two factors which come into play: (i) the
more massive and luminous Es should be better able to hold onto their
dwarf companions, and (ii) the dense environment of the cluster core
creates extreme tidal forces.  This may explain the high fraction of
liberated galaxies, those moving freely in the general potential well
of the cluster.  However small, the number of remaining bound
companions can help us to understand the evolutionary state of the
cluster (Binggeli 1993).  In a study of the small-scale clustering
properties of dwarf galaxies, Vader and Sandage (1991) examined dwarf
galaxies in the vicinity of E and E/S0 type galaxies, and determined
that dwarf galaxy companions are gravitationally bound to distances of
150 kpc.  Ferguson (1992) analyzed the galaxy distribution in the
Virgo cluster, and estimated that on the order of 7 percent of the
galaxy sample are bound companions, located within 80-150 kpc of the
primary.  It is this population of dwarf companions bound to giants
that we attempt to detect in Coma.  Since giants in our region of the
Coma cluster core are typically separated (in projection) by $\lesssim
3$ arcmin ($\lesssim 63h^{-1}$ kpc), we avoid sample contamination due
to crowding by restricting our analysis of dE companions to small
circular regions ($r \simeq 13.9h^{-1}$ kpc) centered on the giants.

As discussed in Section 2.2 and in Secker \& Harris (1996), our dE
galaxy detections are incomplete very near the cluster giants.  To
alleviate this problem, we modeled and subtracted the intensity
profiles of 10 giants, detected and measured anew all objects in these
regions, and then matched object lists to yield a consistent and
complete catalog for these regions.  The isophotal models were
constructed within IRAF using the tasks {\em ellipse} and {\em bmodel}
of the {\em isophote} package within STSDAS.  These tasks implement
the method of Jedrzejewski (1987) for iterative modeling of elliptical
isophotes, and allow for variable center position, ellipticity
$\epsilon$ and position angle $\theta$ at each semimajor axis radius
$r$.  In all cases, we allowed $\epsilon$ and $\theta$ to vary freely,
in order to achieve the best subtraction of the galaxy light (Figure
\ref{ISOmodel}).  In all cases the subtraction was excellent over the
entire radial range, with the exception of the centermost pixels.  We
cross-referenced our positions for these ten galaxies with the
compilation of Kent \& Gunn (1982): Table \ref{tbl4} provides the
NGC/IC names, $(\alpha,\delta)$ and pixel positions, total $R$
magnitudes ($2r_1$ aperture magnitudes), and the radial extent of the
models for the 10 giants considered here.

With the light profiles of the giants subtracted, we proceeded to
detect objects in these areas, using the same analysis methods
described in Paper I: the galaxy-subtracted $R$- and $B$-band master
images were ring median filtered, after which our {\sc DYNAMO}
software (Paper I) was used to detect and measure objects.  Total $R$
magnitudes and $(B-R)$ colors were used to select a color- and
magnitude-restricted subsample of objects (as described in Section
2.4), and objects falling on the bright-stellar sequence were culled
from the list.  The new object lists for the areas around the 10 giant
galaxies was then merged with the final object sample described in
Section 2.3, eliminating any objects which are common to both samples.
Near the center positions of the ten subtracted galaxies, the Poisson
noise inherent in the image is the greatest.  Because of the problem
that this introduces for detection completeness, and because of poor
galaxy subtraction, these small circular regions were excluded from
the analysis.  In addition to these central regions, three of the ten
giants were sufficiently close together that the areas of
consideration around them overlapped; for these cases, wedge-shaped
areas which included the adjacent galaxies were omitted from the
analysis.  In all cases the areas used to compute number densities
were corrected for these excluded areas.
 
Using this spatially complete subsample of objects, we performed
number counts of objects in circular regions of radius 75 pixels
centered on each giant; dividing by the total useable area yielded the
corresponding number densities, $\sigma_{\rm gal}$.  With $N$ and $A$
representing the number of detections and the effective area, we have
$\sigma_{\rm gal} = N/A \pm \sqrt{N}/A$.  For statistical background
correction, we computed a {\em local} background number density,
$\sigma_{\rm bkg}$, near the position of each giant.  By correcting
with $\sigma_{\rm bkg}$, we forgo the need to deproject the 2D galaxy
distribution to a 3D one, since on a statistical basis, any excess of
galaxies above $\sigma_{\rm bkg}$ will be a result of companion
galaxies.  The resulting number densities are provided in Table
\ref{tbl4}.  The calculation for the background-subtracted number of
objects projected within the 75 pixel radius, $N_{75, i}$, around each
giant galaxy, is given by

\begin{equation}
N_{75,i} = A_i \left( \sigma_{{\rm gal},i} - \sigma_{{\rm bkg},i} \right).
\end{equation}
 
\noindent Values of $N_{75,i}$ for our sample of 10 giants are given in
Table \ref{tbl4}.  On average, there are $4\pm1$ objects per giant in
excess of the local background levels, indicating that we have
evidence for a small but significant population of bound companions.
While this result is consistent with the vast majority of dEs being unbound
and free to move in the global cluster potential, it also suggests
that there is a measurable population of bound companions around the
most luminous giants in the cluster core.  A more sophisticated
follow-up analysis would be warranted, to study a wider range of giant
galaxies, with a greater range of position in the cluster and total
magnitude.  In Figure \ref{NvsRmag} we plot $N_{75}$ versus $R$
magnitude for each of the ten giant galaxies.  While there is
significant scatter, there is a weak trend of increasing density with
increasing luminosity of the giant galaxy.  A weighted least-squares
fit yields a best-fit line with a slope 
of $\Delta N_{75}/\Delta R = -5.2\pm2.9$. While this result is
only of slight statistical significance, it works in the sense that
the more luminous and therefore massive galaxies are better able to
bind their companion galaxies, as expected.

\acknowledgments

This paper is based upon thesis research conducted by J.S. while at
McMaster University.  The research was supported in part by: the
Natural Sciences and Engineering Research Council of Canada (through a
grant to W.E.H.), the Department of Physics and Astronomy at McMaster
University, a grant to J.S. from NASA administered by the American
Astronomical Society, a Fullam Award to J.S. from Dudley Observatory,
and the Ontario Ministry of Colleges and Universities (through an
Ontario Graduate Scholarship to J.S).  We would like to thank
P.R. Durrell, D.E. McLaughlin, C.J. Pritchet and the referee, J.
Loveday, for their helpful comments, and to thank S. Holland for his 
help with the observations.

\clearpage

\begin{center}
\begin{deluxetable}{ccccccccc}
\footnotesize
\tablecaption{Calibration summary for April 1991 KPNO 4m run. \label{tbl1}}
\tablewidth{0pt}
\tablehead{\colhead{NIGHT} & \colhead{Total} & \colhead{Filter} & 
\colhead{$ZP_R$} & \colhead{$a_{1}$} & \colhead{$a_{2}$} & \colhead{$ZP_B$} & 
\colhead{$a_{3}$} & \colhead{$a_{4}$} \nl
& \colhead{Standards}  & & & & & & & }
\startdata
1 & 71 & R  &  22.418$\pm$0.016  & 0.130 & --0.002$\pm$0.013 & -- & -- & --\nl
1 & 59 & B  & -- & --& --& 21.081$\pm$0.016 & 0.283 & --0.042$\pm$0.014    \nl
2 & 46 & R  &  22.408$\pm$0.018  & 0.130 & --0.002$\pm$0.013 & -- & -- & --\nl
2 & 34 & B  & -- & --& --& 21.109$\pm$0.018 & 0.283 & --0.042$\pm$0.014    \nl

\enddata
\tablenotetext{1}{Observations made with the TE 2K CCD at PF of the KPNO 4m 
on April 9th and 10th, 1991.} 
\tablenotetext{2}{No errors are provided for $a_{1}$ or $a_{3}$,
average CTIO extinction coefficients from Landolt (1983).}
\end{deluxetable}
\end{center}

\clearpage

\begin{center}
\begin{deluxetable}{cccccccccc}
\footnotesize
\tablecaption{Luminosity Function For Coma Cluster Galaxies \label{tbl2}}
\tablewidth{0pt}
\tablehead{ \colhead{$R$ mag} & \colhead{$N_{\rm Galaxy}$} & 
\colhead{$N_{\rm Background}$} & \colhead{$f(R)$} & \colhead{$N_{\rm Total}$}&
\colhead{$R$ mag} & \colhead{$N_{\rm Galaxy}$} & 
\colhead{$N_{\rm Background}$} & \colhead{$f(R)$} & \colhead{$N_{\rm Total}$}}
\startdata
   12.75 &  ~~1 &    --  &  1.000 &  ~~1$\pm$1~ &    17.75 &  ~30 &  ~~2.6 &  
1.000 &  ~27$\pm$6~ \nl
   13.25 &  ~~6 &    --  &  1.000 &  ~~6$\pm$2~ &    18.25 &  ~49 &  ~~2.6 &  
0.989 &  ~47$\pm$7~ \nl
   13.75 &  ~10 &    --  &  1.000 &  ~10$\pm$3~ &    18.75 &  ~61 &  ~28.3 &  
0.967 &  ~34$\pm$10 \nl
   14.25 &  ~16 &  ~~2.6 &  1.000 &  ~13$\pm$4~ &    19.25 &  ~94 &  ~48.9 &  
0.944 &  ~48$\pm$12 \nl
   14.75 &  ~13 &    --  &  1.000 &  ~13$\pm$4~ &    19.75 &  135 &  ~61.8 &  
0.922 &  ~79$\pm$15 \nl
   15.25 &  ~22 &    --  &  1.000 &  ~22$\pm$5~ &    20.25 &  202 &  131.4 &  
0.900 &  ~79$\pm$19 \nl
   15.75 &  ~~9 &    --  &  1.000 &  ~~9$\pm$3~ &    20.75 &  267 &  164.9 &  
0.878 &  116$\pm$22 \nl
   16.25 &  ~17 &  ~~2.6 &  1.000 &  ~14$\pm$4~ &    21.25 &  346 &  293.7 &  
0.856 &  ~61$\pm$27 \nl
   16.75 &  ~23 &  ~~5.2 &  1.000 &  ~18$\pm$5~ &    21.75 &  481 &  386.4 &  
0.833 &  114$\pm$32 \nl
   17.25 &  ~23 &  ~~5.2 &  1.000 &  ~18$\pm$5~ &    22.25 &  721 &  651.7 &  
0.811 &  ~85$\pm$41 \nl
\enddata
\tablenotetext{1}{The number counts in column 3 includes a correction for
the ratio of areas, $A_p/A_c = 2.576$.}
\tablenotetext{2}{$N_{\rm Total} = INT \left[ f\times(N_{\rm Galaxy}-
N_{\rm Background}) \right]$.}
\end{deluxetable}
\end{center}

\clearpage

\begin{center}
\begin{deluxetable}{ccccccccc}
\footnotesize
\tablecaption{Radial Number Densities For Cluster Galaxies \label{tbl3}}
\tablewidth{0pt}
\tablehead{\colhead{$R_{\rm in}^{'}$} & \colhead{$R_{\rm out}^{'}$} & 
\colhead{${\overline{R}}^{'}$} & \colhead{$N_{\rm bright}$} & \colhead{Area}  
& \colhead{$\sigma_{\rm bright}$} & \colhead{$N_{\rm faint}$} & \colhead{Area}
& \colhead{$\sigma_{\rm faint}$} \nl
&&&& \colhead{arcmin$^2$} & \colhead{$N/$arcmin$^2$} && \colhead{arcmin$^2$}
& \colhead{$N/$arcmin$^2$} }
\startdata
~1.33 & ~3.33 & ~2.10 & 25$\pm$5.0 & ~29.28 & 0.78$\pm$0.17 & 101$\pm$10
& 28.04 & 1.11$\pm$0.39 \nl
~3.33 & ~5.33 & ~4.21 & 29$\pm$5.4 & ~53.91 & 0.47$\pm$0.10 & 205$\pm$15
& 52.77 & 1.40$\pm$0.31 \nl
~5.33 & ~7.33 & ~6.25 & 34$\pm$5.8 & ~76.96 & 0.39$\pm$0.08 & 236$\pm$16
& 73.01 & 0.74$\pm$0.24 \nl
~7.33 & ~9.33 & ~8.27 & 48$\pm$6.9 & 102.60 & 0.41$\pm$0.07 & 365$\pm$20
& 99.20 & 1.19$\pm$0.22 \nl
~9.33 & 11.33 & 10.28 & 40$\pm$6.3 & ~93.03 & 0.36$\pm$0.07 & 312$\pm$18
& 92.50 & 0.88$\pm$0.22 \nl
11.33 & 13.33 & 12.29 & 29$\pm$5.4 & ~77.25 & 0.31$\pm$0.07 & 300$\pm$18
& 76.34 & 1.44$\pm$0.26 \nl
13.33 & 15.33 & 14.29 & 33$\pm$5.7 & ~70.08 & 0.40$\pm$0.08 & 213$\pm$15
& 68.82 & 0.61$\pm$0.24 \nl
15.33 & 17.33 & 16.30 & 13$\pm$3.6 & ~68.83 & 0.12$\pm$0.05 & 236$\pm$16
& 68.77 & 0.94$\pm$0.25 \nl
17.33 & 19.33 & 18.30 & 13$\pm$3.6 & ~58.35 & 0.15$\pm$0.06 & 189$\pm$14
& 57.73 & 0.78$\pm$0.27 \nl
19.33 & 23.33 & 21.23 & 16$\pm$4.0 & ~61.66 & 0.19$\pm$0.07 & 181$\pm$14
& 61.53 & 0.45$\pm$0.25 \nl
\enddata
\tablenotetext{1}{The bright sample includes 280 objects brighter than 
$R = 19$ mag, while the faint sample includes 2246 dE galaxies with
$19.0 < R \leq 22.5$ mag.  Both samples are incomplete within 150 
pixels ($1.325$ arcmin) of NGC 4874 and NGC 4889.  The number counts given in
columns 4 and 7 are corrected for magnitude incompleteness. }
\tablenotetext{2}{The slight differences in annular area between the bright 
and faint samples results from the excluded regions illustrated in 
Figure \ref{SpatialE}(b).}
\tablenotetext{3}{The number densities given in columns 6 and 9 are 
corrected for the mean control field levels: $0.23\pm 0.03$ per 
arcmin$^2$ and $2.33 \pm 0.09$ per arcmin$^2$ respectively.}
\tablenotetext{4}{$\overline{R} = \sqrt{R_{\rm in} R_{\rm out}}$, the 
geometric mean radius.}
\end{deluxetable}
\end{center}

\clearpage

\begin{center}
\begin{deluxetable}{lccccccccc}
\footnotesize
\tablecaption{Background-Corrected Number Counts Centered on Early-type Cluster Giants \label{tbl4}}
\tablewidth{0pt}
\tablehead{
\colhead{NGC/IC}&\colhead{$\alpha$}&\colhead{$\delta$}&\colhead{$X$}&
\colhead{$Y$} &\colhead{$R_{\rm bmodel}$} & \colhead{$R$} &
\colhead{$\sigma_{\rm bkg}$} & \colhead{$\sigma_{\rm gal}$} & \colhead{$N$} \nl
& (2000) & (2000) & \colhead{px}& \colhead{px} & \colhead{px} & \colhead{mag}&
\colhead{$N/$arcmin$^2$} & \colhead{$N/$arcmin$^2$} & }
\startdata
IC 3957 &   12:59:06.9 &  27:46:13.6 & ~342.84 &--288.79 & 58.3 & 14.26 & 
3.5$\pm$0.5 &  3.4$\pm$3.4 & 4.7$\pm$4.6 \nl
IC 3959 &   12:59:08.1 &  27:47:13.6 & ~361.75 &--177.71 & 68.9 & 13.73 & 
3.5$\pm$0.5 &  3.4$\pm$3.4 & 4.7$\pm$4.6 \nl
IC 3963 &   12:59:12.9 &  27:46:37.7 & ~493.90 &--243.61 & 50.4 & 14.23 & 
3.5$\pm$0.5 &  5.5$\pm$3.0 & 7.5$\pm$4.1\nl
IC 3960 &   12:59:08.0 &  27:51:25.6 & ~358.10 & ~305.42 & 39.8 & 14.22 & 
4.4$\pm$0.9 &  1.7$\pm$2.5 & 2.4$\pm$3.4 \nl
NGC 4883&   12:59:55.9 &  28:02:02.5 & 1572.88 & 1519.05 & 37.1 & 13.85 & 
3.6$\pm$0.7 & -0.7$\pm$1.5 &-0.9$\pm$2.1 \nl
IC 4026 &   13:00:21.6 &  28:02:44.9 & 2228.01 & 1598.08 & 37.1 & 14.21 & 
6.0$\pm$0.5 & -1.6$\pm$1.9 &-2.1$\pm$2.6 \nl
IC 4045 &   13:00:48.5 &  28:05:21.5 & 2893.21 & 1892.28 & 37.1 & 13.55 & 
5.3$\pm$0.5 &  0.7$\pm$2.1 & 0.9$\pm$2.9 \nl
NGC 4908&   13:00:51.5 &  28:02:27.5 & 2963.71 & 1565.89 & 42.4 & 13.30 & 
3.3$\pm$0.6 &  8.6$\pm$3.0 &11.8$\pm$4.2\nl
NGC 4906&   13:00:39.6 &  27:55:27.3 & 2664.24 & ~760.03 & 37.1 & 13.89 & 
3.1$\pm$0.3 &  1.4$\pm$1.8 & 1.9$\pm$2.5 \nl
IC 4051 &   13:00:53.3 &  28:00:21.5 & 3035.21 & 1326.23 & 90.1 & 13.05 & 
3.3$\pm$0.7 &  5.6$\pm$2.7 & 7.7$\pm$3.7 \nl
\enddata
\tablenotetext{1}{The magnitudes given in column 5 were derived using $2r_1$ apertures; as noted in Paper I, these underestimate the total magnitude for de Vaucouleurs-profile galaxies such as these.}
\tablenotetext{2}{$\sigma_{\rm gal}$ and $N$ represent the background-corrected number density and number counts within a circle of 75 pixel radius centered on the $(X,Y)$ center of the galaxy.} 
\end{deluxetable}
\end{center}

\clearpage

\figcaption[MATCHEDcompfcn.eps]{The magnitude completeness function was 
derived by comparing photometry for overlapping regions of the program
field. The solid circles represent the completeness function for
objects restricted to the color range $0.7 \le (B-R) \le 1.9$ mag, the
color range of our dE galaxy sample. The open circles represent the
completeness function for all objects, shown here for comparison. The
straight dashed line is our adopted completeness function,
illustrating that our final object lists are 80 percent complete at
the limiting magnitude $R_{2r_1} = 22.5$ mag. \label{MATCHEDcompfcn}}

\figcaption[OVERLAPcompare.eps]{Total magnitude and color comparison 
for objects in common between the {\sc NGC 4874} field and the other
two the program fields.  Crosses denote the 416 objects on the
{\protect {\sc NGC 4889}} field, while the solid circles denote the
448 objects on the {\protect {\sc NGC 4874 South}} field.  (Upper
Panel) The scatter in the magnitude estimates is consistent with
photometric error, yet there is a slight $\simeq 0.05$-mag bias
between total magnitudes measured on the {\protect {\sc NGC 4889}}
field and those measured on the {\protect {\sc NGC 4874}} field.
(Lower panel) The scatter in the $(B-R)$ color estimates is consistent
with photometric error, and there is no evidence for bias between
fields. \label{OVERLAPcompare}}

\figcaption[Fig3a.eps]{512$\times$512 pixel$^2$ sections of the master
$R$-band CCD images, with North upwards and East to the right.  On
each of the images, the most luminous cluster dEs are labeled.  A
fraction of the fainter dE galaxies are obvious as
centrally-concentrated low-surface-brightness objects; however, these
fainter objects also consist of foreground stars and noncluster
galaxies: (a) Northeast of NGC 4889.  The $R$-band total magnitudes
and $(B-R)$ colors for objects 1 through 6 are: (16.82, 1.58); (17.24,
1.27); (17.84, 1.43); (18.49, 1.47); (18.53, 1.38); (18.79, 1.29). (b)
South of NGC 4874.  The $R$-band total magnitudes and $(B-R)$ colors
for objects 1 through 4 are: (17.69, 1.38); (17.86, 1.41); (17.94,
1.41); (19.13, 1.32).  \label{Fig3a}}

\figcaption[FinalCMD.eps]{Composite CMD For The Coma Cluster Core. 
(Left panel) The object lists for the three program fields have been
added together, and the final composite CMD for 3723 objects (2526
within the restricted color range) is illustrated here. (Right panel)
The control field CMD is replotted within the same limits, for
comparison. It consists of a total of 1164 objects, 694 of which are
within the restricted color range. \label{FinalCMD}}

\figcaption[RADmmnt.eps]{Variation of the $r_1$ radius with total $R$ 
magnitude.  The scale radius of the cluster galaxies is traced by the
radial moment $r_1$, and plotted here as a function of the apparent
total $R$ magnitude. We have plotted the color-restricted sample of
program-field objects (left panel) along with the color-restricted
sample of control field objects (right panel); the cluster galaxies
are apparent as an excess of nonstellar objects at all
magnitudes. \label{RADmmnt}}

\figcaption[SRFbright.eps]{The dependence of central surface brightness 
$I_{\rm c}$ as a function of total $R$ magnitude, for the color restricted
sample of program-field objects (left panel) and control-field objects
(right panel). The stellar sequence can be seen to begin at $R = 19.5$ mag,
and the cluster galaxies lie in a diffuse band at lower $I_{\rm c}$.
\label{SRFbright}}

\figcaption[CMDanalysis.eps]{An analysis of the color-magnitude correlation
for the cluster dE galaxies.  (left panel) Reproduced here is the
color-restricted sample of program-field objects, shown as the small
points.  The solid circles represent the median color values (in 1-mag
bins) of these program-field objects, while the open circles indicate
the median color values (in the same 1-mag bins) for the control-field
objects. The solid line is a regression line fit to the five points
shown; the dashed line is an extrapolation of the above line to
fainter magnitudes.  (right panel) The program- and control-field CMDs
were binned, and an area-scaled control-field CMD was subtracted from
the program-field CMD.  There is one vertical line segment per
color-magnitude bin, with a height proportional to the number of
objects in the bin.  The bins have a height of 1.0 mag, and a width of
0.1 mag.  The maximum number of objects in a bin is 50, while several
bins contain a negative number of objects.  The solid circles
correspond to the mean galaxy color (an average color weighted by the
number of objects per bin) in a 1-mag bin. The regression line from
the left panel provides an excellent fit over the entire magnitude
range.  This fiducial dE sequence indicates that a one magnitude
brightening in $R$ corresponds to a redder color by $(B-R) = 0.056$
mag.  \label{CMDanalysis}}

\figcaption[FaintendLF.eps]{The faint-end of the completeness-corrected
and background-subtracted luminosity function for our sample of Coma
cluster galaxies, as defined in Table \ref{tbl2}. The solid line is a
weighted least-squares fit over the range shown, which yields a slope of
$\alpha = -1.41\pm0.05$ for the Schechter luminosity function. 
\label{FaintendLF}}

\figcaption[ModelLF.eps]{The composite galaxy luminosity function (LF) 
for the Coma cluster core was decomposed into the contribution from
the giants (i.e., a log-normal distribution) and from the dwarf
ellipticals (i.e., a Schechter function).  The models have a Gaussian
dispersion which varies between $\sigma = 0.8$ mag and $\sigma = 1.2$
mag; Three of these models are illustrated in this plot.  The best-fit
model has $\sigma = 0.80$ mag, and for this case we illustrate the
decomposition of the model into its two components (dotted lines).
Note that the three faintest points were not included in the fit, as
they adversely affect normalization over the region of
interest. \label{ModelLF}}

\figcaption[SpatialE.eps]{The spatial distribution of candidate cluster 
members.  (a) Program-field objects brighter than $R = 19$ mag (mostly
cluster Es and dEs), show a strong clustering around NGC 4874 and NGC
4889, located in this plot at (1061,1010) and (1875,1124)
respectively.  The small dotted circles denote incomplete regions; (b)
Program-field objects fainter than $R = 19$ mag (faint cluster dEs and
a uniformly distributed background population).  The large dotted
circles are centered on NGC 4874, and denote the annular boundaries
used for computing the radial surface density profiles. The small
solid circles denote incomplete areas, corresponding to NGC 4874, NGC
4889, a bright star, and several other extended galaxies. \label{SpatialE}}

\figcaption[Artstarcomp.eps]{The detection completeness functions 
derived from artificial star simulations.  The four radial zones are
numbered such that the innermost zone around NGC 4874 is Zone 1.  The
solid circles and associated uncertainties were computed knowing the
number of artificial stars input in each magnitude bin, and from the
number of these objects which were subsequently detected.  The dotted
lines illustrate least-squares fits of an appropriate interpolation
function. \label{Artstarcomp}}

\figcaption[RDFfinal.eps]{Radial number density profile log($N$/arcmin$^2$) 
versus log($R_{\rm cc}$/arcmin) for Coma cluster galaxies, plotted
over the clustercentric range $1.33 \leq R_{\rm cc} \leq 23.33$
arcmin.  The top panel correspond to the sample of cluster E and
bright dEs, and the bottom panel corresponds to the sample of faint
dEs.  while the solid line gives the best-fit King model.  The solid
line illustrates the best-fit King model, which in both cases provides
an excellent fit to the observed distributions.\label{RDFfinal}}

\figcaption[ISOmodel.eps]{Original and model-subtracted images of five 
cluster giants. The upper left section of the image illustrates the
cluster giants IC 3957, IC 3959 and IC 3963 (Table \ref{tbl4}), while
the upper right shows the image after isophotal modeling and
subtraction.  The lower left image shows NGC 4908 and IC 4051; the
lower right image shows the image after these two galaxies have been
subtracted.  In all cases the isophotal models subtracted very well,
except for small regions at the galaxy centers.
\label{ISOmodel}}

\figcaption{The background-subtracted number of objects within a 75 pixel
radius around each of the ten giant Es in the cluster core is plotted
versus an estimate of the total $R$-band magnitude of the giants.  The
weighted least-squares regression line is also shown, though it has
low statistical significance.
\label{NvsRmag}}

\end{document}